\documentclass[review]{article}
\usepackage{jheppub0}
\usepackage{graphicx,color}
\usepackage{amsmath, amssymb, bm}
\numberwithin{equation}{section}
\newcommand{\lambdabar}{\lambda\kern -0.5em\raise 0.5ex \hbox{--}}

\begin{document}


\title{Ladder Operators in Repulsive Harmonic Oscillator with Application to the Schwinger Effect}

\author[a]{Kenichi Aouda,}
\author[a]{Naohiro Kanda,}
\author[a]{Shigefumi Naka,}
\affiliation[a]{Department of Physics, College of Science and Technology, Nihon University, Tokyo 101-8308, Japan}
\author[b]{and Haruki Toyoda}
\affiliation[b]{Junior College, Funabashi Campus, Nihon University, Funabashi 274-8501, Japan}

\emailAdd{aouda@phys.cst.nihon-u.ac.jp}
\emailAdd{nkanda@phys.cst.nihon-u.ac.jp}
\emailAdd{naka@phys.cst.nihon-u.ac.jp}
\emailAdd{toyoda.haruki@nihon-u.ac.jp}

\abstract{ \input{abstract} }

\abstract{
The ladder operators in harmonic oscillator are a well-known strong tool for various problems in physics. In the same sense, it is sometimes expected to handle the problems of repulsive harmonic oscillator in a similar way to the ladder operators in harmonic oscillators, though their analytic solutions are well known. In this paper, we discuss a simple algebraic way to introduce the ladder operators of the repulsive harmonic oscillators, which can reproduce well-known analytic solutions.  Applying this formalism, we discuss the charged particles in a constant electric field in relation to the Schwinger effect; the discussion is also made on a supersymmetric extension of this formalism. 
}



\maketitle

\section{Introduction}

The algebraic approaches to the potential problems in quantum mechanics are commonly used ways from the early state of those fields\cite{Dirac}. In particular, the harmonic oscillators (HOs) give a good operative example of an algebraic approach to the eigenvalue problems in terms of the ladder operators, the annihilation and creation operators $(\hat{a},\hat{a}^\dag)$ characterized by $[\hat{a}, \hat{a}^\dag]=1$. In such a dynamical system, the eigenvalue problem of Hamiltonian can be solved exactly by use of those ladder operators without depending on the representation of the eigenstates\cite{Dirac,Landau} and, if we take the coordinate representation of those states, the eigenstates will be reduced to the well-known analytic solutions expressed in terms of Hermite polynomials. The use of ladder operators also provides necessary tools in the field theories, since the dynamical degrees of freedom of bosonic-free fields are decomposed into those of infinite harmonic oscillators.

In comparison with HOs, the physical applications of the repulsive harmonic oscillators (RHOs)
\footnote{The {\it inverted oscillator} or {\it reversed oscillator}, in other words.}
 are limited, since the Hamiltonian of RHOs is parabolic and its eigenstates are scattering states. The algebraic approaches to RHOs, however, have been tried from a few different viewpoints: the dynamical groups including RHOs \cite{Kalnins, Wolf}, the analytic continuation of angular velocity $\omega\rightarrow \pm i\omega$ in HOs\cite{Barton, Shimbori,Rajeev}, the Bose systems in SUSY quantum mechanics\cite{Mota,Shimbori-SUSY}, and so on.

On the other hand, it is known that the eigenvalue problems of the RHO Hamiltonian are reduced to solve Weber's equation, which has analytic solutions so-called parabolic cylinder functions or the Weber functions\cite{Whittaker,Hara}. The relation between the algebraic approaches to RHOs and the analytic solutions, however, is not always clear. It is also important to study the completeness of the states constructed out of the algebraic approaches, since the trace calculations in physical applications require such a property of those states. The purpose of this paper
\footnote{
This paper is a supplemented version of the paper with doi:\,10.1103/PhysRevD.102.025002.}
 is, thus, to give a simple algebraic approach to the eigenvalue problems of RHOs by introducing Hermitian ladder operators $(A,\bar{A})$ characterized by $[A,\bar{A}]=i$. 

We can show that the dynamical variables of RHOs can be represented in the functional spaces constructed out of $(A,\bar{A})$ with two cyclic states $(\phi_0,\bar{\phi}_0)$ satisfying $A\phi_0=\bar{A}\bar{\phi}_0=0$\cite{Bermudez}. Here, the $\{\bar{A}^n{\phi}_0\}$ and $\{A^n\bar{\phi}_0\}\,(n\in \mathbb{N})$ are conjugate, complex conjugate in the $x$-representation, states which form orthonormal pairs, though those themselves are not square integrable. Those pairs become complex conjyugate of each other in $x$ representation. As the result, those states form a discrete basis of a space of functionals $\bm{\Phi}^{\times}$, which includes the Hilbert space $\mathcal{H}$ for the RHO. It is also shown that there exist continuous bases $\{\phi_\sigma,\bar{\phi}_\sigma\}\, (\sigma\in \mathbb{R})$ in $\bm{\Phi}^{\times}$, which are respective eigenstates of $A$ and $\bar{A}$.

In the next section, we study those continuous and discrete bases given in terms of the ladder operators with their cyclic states. In that place, the completeness of those bases is discussed carefully. The discussions are also made on the eigenvalue problems of the RHO Hamiltonian by considering the relation between the ladder operator formalism and the well-known analytic solutions.

In Sec. III, we discuss the applications of the present ladder operator formalism to two topics: one is a problem of charged particles under a constant electric field, the problem of the Schwinger effect\cite{Schwinger}.  This dynamical system is equivalent to RHO and the discrete basis in the ladder operator formalism is shown to be useful to evaluate that effect. As another topic, we study an extension of RHOs to a model of supersymmetry (SUSY) quantum mechanics by taking the advantage of the ladder operator formalism, though such an extension has been discussed from the early stages of RHOs. We focus our attention on the fact that the Schwinger effect for fermions is closely related to such an extended model.

Section IV is devoted to the summary of our results. In the Appendices, some mathematical problems used in the text are discussed: the analytic solutions of Hamiltonian eigenstates, a proof of completeness, and the evaluation of the Schwinger effect for fermions.

\section{Ladder operators in repulsive harmonic oscillators}

\subsection{Summary of standard harmonic oscillators} 

To begin with, we summarize the ladder operator approach to the problems of the usual harmonic oscillator, to which the Hamiltonian operator of a mass $m$ particle with the characteristic frequency $\omega$ of the oscillation in one-dimensional space is given by
\begin{align} 
 \hat{H} &\!=\!\frac{1}{2m}\hat{p}^2+\frac{m\omega^2}{2}\hat{x}^2\!=\!\frac{\hbar\omega}{2}\left(\hat{a}^\dag\hat{a}+\hat{a}\hat{a}^\dag \right)\!=\!\hbar\omega\left(\hat{N}+\frac{1}{2}\right), \label{h.o}
\end{align}
where $\hat{N}=\hat{a}^\dag\hat{a}$ and
\begin{align}
\begin{split}
 \hat{a} &=\sqrt{\frac{m\omega}{2\hbar}}\hat{x}+\frac{i}{\sqrt{2m\hbar\omega}}\hat{p} \,, \\
 \hat{a}^\dag &=\sqrt{\frac{m\omega}{2\hbar}}\hat{x}-\frac{i}{\sqrt{2m\hbar\omega}}\hat{p}\, .
\end{split}
\end{align}
By definition $\langle\Phi|\hat{N}|\Phi\rangle=\|\hat{a}\Phi\|^2\geq 0$; then, because of $[\hat{a},\hat{a}^\dag]=1$, one can verify that $[\hat{H},\hat{a}^\dag]=\hbar\omega\hat{a}^\dag,\,[\hat{H},\hat{a}]=-\hbar\omega\hat{a}$, and $\|\hat{H}\Phi\|\geq \frac{\hbar\omega}{2}$
 on a state $\Phi$ normalized so that $\|\Phi\|^2=1$ \footnote{$\|\hat{H}\Phi\|^2=\langle\Phi|\hat{H}^2|\Phi\rangle\geq \langle\Phi|\hat{H}|\Phi\rangle^2=(\hbar\omega)^2\left(\langle\Phi|\hat{N}|\Phi\rangle +\frac{1}{2}\right)^2.$} . This means that starting from the ground state $\Phi_0$ defined by $\hat{a}\Phi_0=0$ with $\|\Phi_0\|^2=1$, the states
\begin{align}
 \Phi_n=\frac{1}{\sqrt{n!}}\hat{a}^{\dag n}\Phi_0 ~~~ (n=0,1,2,3,\cdots)
\end{align}
satisfy the eigenvalue equations
\begin{align}
\hat{H}\Phi_n=\hbar\omega\left(n+\frac{1}{2}\right)\Phi_n ~~~ (n=0,1,2,3,\cdots), \label{h.o.energy}
\end{align}
and the normalization $\langle\Phi_n|\Phi_m\rangle=\delta_{n,m}$. The importance is that the states $\{\Phi_n\}$ really form a complete basis of the functional space $V$, in which the canonical operators $(\hat{x},\hat{p})$ are represented. Namely, in terms of the bra and the ket states, the operator
\begin{align}
 \hat{I}=\sum_{n=0}^\infty |\Phi_n\rangle\langle \Phi_n| \label{unit-1}
\end{align}
is the unit operator in the functional space $V$ and one can verify
\begin{align}
 \langle x|\hat{I}|x^\prime\rangle=\delta(x-x^\prime), \label{unit-2}
\end{align}
where $\{|x\rangle\}$ are the eigenstates of $\hat{x}$ characterized by $\hat{x}|x\rangle=x|x\rangle$ and $\langle x|x^\prime\rangle=\delta(x-x^\prime),\,(x,x^\prime\in R)$. Furthermore, if it is necessary, the $x$ representation of $\Phi_n$ can be written explicitly in terms of the Hermitian polynomial $H_n(x)$ so that $\Phi_n(x)=\langle x|\Phi_n\rangle =\sqrt{\frac{1}{2^n n!}\sqrt{\frac{m\omega}{\pi\hbar}}}e^{-m\omega x^2/2\hbar}H_n(x\sqrt{m\omega/\hbar})$.
 
\subsection{The case of repulsive harmonic oscillators}

Now, for a repulsive harmonic oscillator, the Hamiltonian operator $\hat{H}_r$ is given from $\hat{H}$ in Eq.(\ref{h.o}) by changing the sign of $\frac{m\omega^2}{2}\hat{x}^2$ ; and, a complete basis in the same functional space $V_r$ by means of new ladder operators   can be constructed in roughly parallel with Eqs.(\ref{h.o})-(\ref{unit-2}). Namely, one can start with the expression
\begin{align}
  \hat{H}_r &=\frac{1}{2m}\hat{p}^2-\frac{m\omega^2}{2}\hat{x}^2=-\frac{\hbar\omega}{2}\left(\bar{A}A+A\bar{A}\right), \label{r.h.o}
\end{align}
where
\begin{align}
\begin{split}
 A &=\sqrt{\frac{m\omega}{2\hbar}}\hat{x}-\frac{1}{\sqrt{2m\hbar\omega}}\hat{p} \,, \\
 \bar{A} &=\sqrt{\frac{m\omega}{2\hbar}}\hat{x}+\frac{1}{\sqrt{2m\hbar\omega}}\hat{p}\, .
\end{split}
\end{align}
By definition, $A$ and $\bar{A}\,(\neq A^\dag)$ are Hermitian operators themselves; however, they satisfy a similar algebra as that of $(\hat{a},\hat{a}^\dag)$ such as $[A,\bar{A}]=-[\bar{A},A]=i$. 
Further, in terms of $(A,\bar{A})$, the Hamiltonian operator $\hat{H}_r$ can be written as
\footnote{
In terms of the ladder operator $(\hat{a},\hat{a}^\dag)$ defined in Eq.(\ref{h.o}), the Hamiltonian operator (\ref{r.h.o}) can be represented as   $\hat{H}_r=-\frac{\hbar\omega}{2}(\hat{a}^{\dag 2}+\hat{a}^2)$. From this expression, carrying out the successive canonical ($\neq$ unitary) transformations by $U_1=e^{\frac{i}{2}\hat{a}^{\dag 2}}$ and $U_2=e^{-\frac{i}{4}\hat{a}^2}$, one can find the relation between $\hat{H}_r$ and $\hat{H}$ such that  $U_2U_1\hat{H}_rU_1^{-1}U_2^{-1}=i\hat{H}$. The eigenvalue problem of $\hat{H}_r$ , thus, can also be solved in terms of $(\hat{a},\hat{a}^\dag)$ and these canonical transformations.}
\begin{align}
 \hat{H}_r &=-i\hbar\omega\left(\Lambda+\frac{1}{2}\right)=-i\hbar\omega\left(\bar{\Lambda}-\frac{1}{2}\right), \label{Hamiltonian}
\end{align}
where
\begin{align}
 \Lambda &= -i\bar{A}A \, ~~~ \mbox{and}~~~\bar{\Lambda}=-iA\bar{A}~(\, =\Lambda+1 \,)\,.
\end{align}
Since $\Lambda^\dag=-\Lambda-1\,(\bar{\Lambda}^\dag=-\bar{\Lambda}+1)$,  the Hermiticity of the operator $\hat{H}_r$ given in Eq.(\ref{Hamiltonian}) is formally guaranteed. The eigenvalue problem of $\hat{H}_r$ is, thus, reduced to those of the operators $\Lambda$ and $\bar{\Lambda}\,(\neq \Lambda^\dag)$, which are commutable with each other. 

In order to solve the eigenvalue problem of $\Lambda$ and $\bar{\Lambda}$, let us introduce eigenstates $(\phi_\sigma,\bar{\phi}_\sigma)$ defined by 
\begin{align}
 A\phi_\sigma &=\left(\sqrt{\frac{m\omega}{2\hbar}}\hat{x}-\frac{1}{\sqrt{2m\hbar\omega}}\hat{p}\right)\phi_\sigma=\sigma\phi_\sigma , \label{A-eigen} \\
\bar{A}\bar{\phi}_\sigma &=\left(\sqrt{\frac{m\omega}{2\hbar}}\hat{x}+\frac{1}{\sqrt{2m\hbar\omega}}\hat{p}\right)\bar{\phi}_\sigma=\sigma\bar{\phi}_\sigma , \label{barA-eigen} 
\end{align}
where the $\sigma$ is a real parameter.  Then, the particular states $(\phi_0,\bar{\phi}_0)$ defined by $A\phi_0=\bar{A}\bar{\phi}_0=0$ should be regarded as the counterparts of $\Phi_0$ in the HO. It should be noticed that in spite of the similarity of Eq.(\ref{A-eigen}) to the coherent state equation in the HO, the index $\sigma$ of $\phi_\sigma$ runs over the real continuous spectrum due to the Hermiticity of $A$ and the same is true for $\bar{\phi}_\sigma$.

In the $x$ representation,Eqs.(\ref{A-eigen}) and (\ref{barA-eigen}) can be solved explicitly, and we obtain
\begin{align}
 \phi_{\sigma}(x) &=\sqrt[4]{\frac{m\omega}{2\hbar\pi^2}} e^{i\frac{m\omega}{2\hbar}x^2-i\sqrt{\frac{2m\omega}{\hbar}}\sigma x} , \label{phi-sigma} \\
 \bar{\phi}_{\sigma}(x) &=\sqrt[4]{\frac{m\omega}{2\hbar\pi^2}} e^{-i\frac{m\omega}{2\hbar}x^2+i\sqrt{\frac{2m\omega}{\hbar}}\sigma x} , \label{barphi-sigma} 
\end{align}
where the normalizations of those states are $\langle\phi_\sigma|\phi_{\sigma^\prime}\rangle=\langle\bar{\phi}_\sigma|\bar{\phi}_{\sigma^\prime}\rangle
=\delta(\sigma-\sigma^\prime) $. In this $x$ representation, because of $\bar{\phi}_\sigma(x)=\langle x|\bar{\phi}_\sigma\rangle=\phi_\sigma(x)^*=\langle x|\phi_\sigma^*\rangle$, the bar becomes simply complex conjugation, and the functional space of $\{\phi_\sigma\}$ coincides with that of $\{\bar{\phi}_\sigma\}$ in the aggregate, though $\phi_\sigma$ and $\bar{\phi}_\sigma$ are independent states. Further, one can find the completeness of $(\phi_\sigma,\bar{\phi}_\sigma)$ in the form
\begin{align}
\int d\sigma\langle x|\phi_\sigma\rangle\langle\phi_\sigma|x^\prime\rangle=\int d\sigma\langle x|\bar{\phi}_\sigma\rangle\langle\bar{\phi}_\sigma|x^\prime\rangle=\delta(x-x^\prime).
\end{align}
Thus, the states $\{\phi_\sigma\}$ and their conjugate $\{\bar{\phi}_\sigma\}$ are continuous complete bases
\footnote{Because of $U_A(\hat{x},\hat{p})U_A^{-1}=\left(\sqrt{\frac{\hbar}{m\omega}}A,\sqrt{\hbar m\omega}\bar{A}\right)$ with $U_A=e^{-i\frac{\pi}{8}(A^2+\bar{A}^2)}$, the states $|\phi_\sigma\rangle$ and $|\bar{\phi}_\sigma\rangle$ are unitary equivalents to $|x=\sqrt{\frac{\hbar}{m\omega}}\sigma\rangle$ and $|p=\sqrt{\hbar m\omega}\sigma\rangle$ respectively.}
 of the functional space $\bm{\Phi}^{\times}$, which includes the Hilbert space $\mathcal{H}$ for the RHO in the framework of the Rigged Hilbert space.
\footnote{For the quantum mechanics dealing with continuous spectrum, the rigged Hilbert space\cite{RHS-1,RHS-2} $\bm{\Phi} \subset \mathcal{H} \subset \bm{\Phi}^{\times}$ is useful to include continuous bases in the framework. Here, $\mathcal{H}$ is the Hilbert space with a countable orthonormal basis such as the $\{\Phi_n\}$ in HOs. The $\bf{\Phi}$ is a dense subspace of $\mathcal{H}$ associated with a topology finer than that of $\mathcal{H}$: and the $\bm{\Phi}^{\times}$ is the dual space of $\bm{\Phi}$. The $\{|\phi_\sigma\rangle\},\{|\bar{\phi}_\sigma\rangle\}$, and $\{|x\rangle\}$ are continuous bases of $\bm{\Phi}^{\times}$.}

\begin{figure}
\center
 \includegraphics[width=4cm]{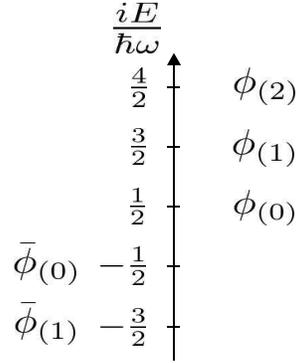}
\caption{There are many types of complete bases in the representation space of $\hat{H}_r$. On the discrete bases $(\phi_{(n)},\bar{\phi}_{(n)})$, the $\hat{H}_r$ takes the eigenvalues shown in the figure on the left of vertical axis.  }
\label{H_r eigenstates}
\end{figure}

In those continuous complete bases $\{\phi_\sigma\}$ and $\{\bar{\phi}_\sigma\}$, the aspect of the states $(\phi_0,\bar{\phi}_0)\in \bm{\Phi}^{\times}$satisfying $A\phi_0=\bar{A}\bar{\phi}_0=0$ are characteristic. First, the $(\phi_0,\bar{\phi}_0)$ should be regarded as the counterparts of the ground state $\Phi_0$ in the HO. Second, those states become cyclic states of $\bm{\Phi}^{\times}$ in the following sense: Writing $(\phi_{(0)},\bar{\phi}_{(0)})=(\phi_0,\bar{\phi}_0)$, one can verify that the states defined by
\begin{align}
 \phi_{(n)}=\bar{A}^n\phi_{(0)}~~\left(\,\bar{\phi}_{(n)}=A^n \bar{\phi}_{(0)} \,\right)\,,~(n=0,1,2,\cdots) 
\end{align}
satisfy the eigenvalue equations
\begin{align}
  \Lambda\phi_{(n)}=n\phi_{(n)}~~\left(\, \bar{\Lambda}\bar{\phi}_{(n)}=-n\bar{\phi}_{(n)} \,\right)\,,~(n=0,1,2,\cdots) .
\end{align}
Namely, on the states $(\phi_{(n)},\bar{\phi}_{(n)})$, the Hamiltonian operator   $\hat{H}_r$ takes discrete eigenvalues (Fig.\,\ref{H_r eigenstates}) such that
\begin{align}
 \left . \begin{matrix}
 \hat{H}_r \phi_{(n)}=-i\hbar\omega\left(n+\frac{1}{2}\right)\phi_{(n)} \rule[-4mm]{0pt}{0mm}\\
 \hat{H}_r\bar{\phi}_{(n)}=i\hbar\omega\left(n+\frac{1}{2}\right)\bar{\phi}_{(n)}
\end{matrix}
 \right\} \,,~(n=0,1,2,\cdots) ,  \label{H-spectra}
\end{align}
which means that there are no ground states for $\hat{H}_r$  as expected from its nonpositive structure.

Those $\{\phi_{(n)},\bar{\phi}_{(n)}\}$ are the generalized eigenstates belonging to $\bm{\Phi}^{\times}$ instead of the Hilbert space for the  RHO. What is important is that the states $\{\phi_{(n)}\}$ and their conjugate $\{\bar{\phi}_{(n)}\}$ are orthogonal each other under the inner  product, which can be determined from the algebra of $(A,\bar{A})$ and the normalization $\langle\bar{\phi}_{(0)}|\phi_{(0)}\rangle\equiv N_0=\sqrt{\frac{i}{2\pi}}$ only. Indeed for $m=n+l\,(l>0)$, one can verify
\begin{align}
 \langle\bar{\phi}_{(m)}|\phi_{(n)}\rangle &=\langle\bar{\phi}_{(0)}|A^lA^n\bar{A}^n|\phi_{(0)}\rangle \nonumber \\
 &=in\langle\bar{\phi}_{(0)}|A^lA^{n-1}\bar{A}^{n-1}|\phi_{(0)}\rangle \nonumber \\
 &=\cdots=i^nn!\langle\bar{\phi}_{(0)}|A^l|\phi_{(0)}\rangle,
\end{align}
which leads to $\langle\bar{\phi}_{(m)}|\phi_{(n)}\rangle=0\,(m>n)$; the same is true for the case $m<n$. Thus, the inner products between any $m,n$ states can be written as
\begin{align}
 \langle\bar{\phi}_{(m)}|\phi_{(n)}\rangle=\delta_{m,n}N_n~~(\,N_n\equiv i^nn!N_0\,),
 \label{normalization}
\end{align}
which gives the meaning of $\{\bar{\phi}_{(n)}\}$ without depending on the representation. Here, the complexity of $N_n$'s again implies that the $\{\phi_{(n)},\bar{\phi}_{(n)}\}$ are not bases in a Hilbert space in spite of the resemblance between those states and $\{\Phi_n\}$ in the HO.

Nevertheless, Eq.(\ref{normalization}) suggests that the operator
\begin{align}
 \hat{I}_r \equiv \sum_{n=0}^\infty \frac{1}{N_n}|\phi_{(n)}\rangle\langle\bar{\phi}_{(n)}|
 \label{unit operator}
\end{align}
plays the role of a unit operator in $\{\phi_{(n)}\}$ space. The expectation $\hat{I}_r=1$, can be confirmed through the equation
\begin{align} A\hat{I}_r &=\sum_{n=1}^\infty\frac{in}{N_n}|\phi_{(n-1)}\rangle\langle\bar{\phi}_{(n)}|=\sum_{n=0}^\infty\frac{i(n+1)}{N_{n+1}}|\phi_{(n)}\rangle\langle\bar{\phi}_{(n+1)}| \nonumber \\
 &=\sum_{n=0}^\infty\frac{1}{N_n}|\phi_{(n)}\rangle\langle\bar{\phi}_{(n)}|A=\hat{I}_rA,
\end{align} 
which can be verified using $\frac{i(n+1)}{N_{n+1}}=\frac{1}{N_n}$. In a similar way, one can derive $\bar{A}\hat{I}_r=\hat{I}_r\bar{A}$. Since $A$ and $\bar{A}$ are composing elements of dynamical variables in RHOs, one can say $\hat{I}_r=c1,\,(c=\mbox{const})$ in the sense of Schur's lemma. Here, the constant in the right-hand side  is necessary to be $c=1$ because of $\hat{I}_r|\phi_{(0)}\rangle=|\phi_{(0)}\rangle$ by Eq.(\ref{normalization}). In Appendix B, we will show directly
\begin{align}
 \langle x|\hat{I}_r|x^\prime\rangle=\delta(x-x^\prime), \label{complete}
\end{align}
which says that the imaginary parts of each term in the right-hand side of  Eq.(\ref{complete}) are cancelled out by the summation with respect to $n$. Thus, by taking into account $(|\phi_{(n)}\rangle\langle\bar{\phi}_{(n)}|)^\dag=|\bar{\phi}_{(n)}\rangle\langle\phi_{(n)}|$ and $1^\dag=1$, Eq.(\ref{unit operator}) is equivalently represented as
\begin{align}
 \hat{I}_r = \sum_{n=0}^\infty \frac{1}{N^*_n}|\bar{\phi}_{(n)}\rangle\langle\phi_{(n)}| \,,
 \label{unit operator-2}
\end{align}
from which one can write the spectral decomposition of $\hat{H}_r$ so that
\begin{align}
 \hat{H}_r &=\sum_{n=0}^\infty \frac{-i\hbar\omega\left(n+\frac{1}{2}\right)}{N_n}|\phi_{(n)}\rangle\langle\bar{\phi}_{(n)}| \label{spectral-1} \\
  &=\sum_{n=0}^\infty \frac{i\hbar\omega\left(n+\frac{1}{2}\right)}{N^*_n}|\bar{\phi}_{(n)}\rangle\langle\phi_{(n)}| \,. \label{spectral-2}
\end{align}
The resultant equations, (\ref{unit operator-2})-(\ref{spectral-2}), also have the meaning independent of the representation equation (\ref{unit operator}). Since  $\hat{H}_r^\dag=\hat{H}_r$, two types of spectral decomposition (\ref{spectral-1}) and (\ref{spectral-2}) are consistent and $(\phi_{(0)},\bar{\phi}_{(0)})$ are not ground states corresponding to any lower bounds of $\hat{H}_r$ but rather, to the cyclic states of $\bm{\Phi}^{\times}$.

The states $\{\phi_{(n)},\bar{\phi}_{(n)}\}$ form a discrete basis of $\bm{\Phi}^{\times}$ in pairs in addition to that those are generalized eigenstates of $\hat{H}_r$. The eigenstates of $\hat{H}_r$ are not limited to those states; we emphasize that the discrete basis $\{\phi_{(n)},\bar{\phi}_{(n)}\}$ is closely related to Weber's functions, which are continuous eigenvalue solutions for an eigenvalue equation of $\hat{H}_r$, by means of the analytic continuation with respect to $n$. In order to verify this, we take notice the formula for a complex $\lambda$:
\begin{align}
\bar{A}^\lambda &=\frac{1}{\Gamma(-\lambda)}\int_0^\infty dt e^{-\bar{A}t}t^{-(\lambda+1)}    \label{Gamma} \\
 &=\frac{1}{\Gamma(-\lambda)}\int_0^\infty dtt^{-(\lambda+1)}e^{-\frac{i}{4}t^2}e^{-t\sqrt{\frac{m\omega}{2\hbar}}\hat{x}}e^{-t\frac{1}{\sqrt{2m\hbar\omega}}\hat{p}}. \label{A.C.}
\end{align} 
Here, Eq.(\ref{Gamma}) seems to hold on the states such as $\{\bar{\phi}_\sigma;\sigma>0\}$, on which $\bar{A}$ becomes an operator with positive eigenvalues. Applying Eq.(\ref{A.C.}) to $\phi_{(0)}(x)$, such a constraint will fade away in the sense of analytic continuation; and, we obtain the expression
\begin{align}
\bar{A}^\lambda\phi_{(0)}(x) &=\sqrt[4]{\frac{m\omega}{2\hbar\pi^2}}\frac{1}{\Gamma(-\lambda)}\int_0^\infty dt \nonumber \\
 &\times t^{-(\lambda+1)}e^{-\frac{i}{4}t^2}e^{-t\sqrt{\frac{m\omega}{2\hbar}}x}e^{i\frac{m\omega}{2\hbar}\left(x+it\sqrt{\frac{\hbar}{2m\omega}}\right)^2} \nonumber \\
&=e^{\frac{i\pi}{4}\lambda}\sqrt[4]{\frac{m\omega}{2\hbar\pi^2}}\frac{e^{-\frac{1}{4}\left(e^{-\frac{i\pi}{4}}\sqrt{\frac{2m\omega}{\hbar}}x\right)^2}}{\Gamma(-\lambda)}\int_0^\infty d\bar{t} \nonumber \\
 &\times \bar{t}^{-(\lambda+1)}e^{-\frac{1}{2}\bar{t}^2}e^{-\bar{t}\left(e^{-\frac{i\pi}{4}}\sqrt{\frac{2m\omega}{\hbar}}x\right)} \nonumber \\
&=e^{\frac{i\pi}{4}\lambda}\sqrt[4]{\frac{m\omega}{2\hbar\pi^2}}D_\lambda\left(z\right)\,, \label{D-function}
\end{align} 
where $\bar{t}=e^{\frac{i\pi}{4}}t$ and $z=e^{-\frac{i\pi}{4}}\sqrt{\frac{2m\omega}{\hbar}}x$. The last equality in Eq.(\ref{D-function}) shows the relationship\cite{Bateman} between $\bar{A}^{\lambda}\phi_{(0)}(x)$ and Weber's function $D_\lambda(z)$ (Appendix A). In a similar manner, one can verify that
\begin{align}
A^\rho\bar{\phi}_{(0)}(x) &=e^{-\frac{i\pi}{4}\rho}\sqrt[4]{\frac{m\omega}{2\hbar\pi^2}}D_\rho(iz), \label{D-function-2}
\end{align}
which can be regarded as the analytic continuation of the relation $\bar{\phi}_{(n)}(x)=\phi_{(n)}^*(x)$ with respect to $n$.  We note that if the $\lambda$ in Eq.(\ref{D-function}) and the $\rho$ in Eq.(\ref{D-function-2}) give the same eigenvalue of $\frac{i\hat{H}_r}{\hbar\omega}$, then $\lambda+\frac{1}{2}=-(\rho+\frac{1}{2})$ or $\rho=-(\lambda+1)$. Therefore, $D_\lambda(z)$ and $D_{-(\lambda+1)}(iz)$ are independent eigenstates of $\frac{i\hat{H}_r}{\hbar\omega}$ belonging to the same eigenvalue $\lambda+\frac{1}{2}$.  This is a well-known result of discrete eigenstates in the eigenvalue problem of RHOs \cite{Whittaker,Hara}. In terms of Weber's $D$ function, the completeness condition (\ref{complete}) can also be represented as
\begin{align}
  \langle x|\hat{I}_r|x^\prime\rangle &=\sum_{n=0}^{\infty} \frac{1}{N_n}\phi_{(n)}(x)\bar{\phi}_{(n)}(x^\prime)^*  \nonumber \\
 &=\sum_{n=0}^\infty \frac{i^n}{N_n}\left(\frac{m\omega}{2\hbar\pi^2}\right)^{\frac{1}{2}}D_n(z)D_n(iz^\prime)^* .
\end{align}

In summary,  the complete bases $\{\phi_\sigma(x)\}$ and $\{\bar{\phi}_\sigma(x)\}$ are respective eigenstates of $A$ and $\bar{A}$  belonging to continuous eigenvalues $\{\sigma\in \mathbb{R}\}$, but those are not eigenstates of $\hat{H}_r$. On the other hand,  the eigenstates $\{\phi_{(n)}(x),\bar{\phi}_{(n)}(x)\}$ are eigenstates of $\hat{H}_r$ with discrete eigenvalues corresponding to the analytic continuation $\omega\rightarrow \pm i\omega$ of the eigenvalues in Eq.(\ref{h.o.energy}). The $\{\phi_{(n)}(x)\,\bar{\phi}_{(n)}(x)\}$ form a discrete basis of $\bm{\Phi}^{\times}$ in pairs.  The $\{D_\lambda(z);\,\lambda\in \mathbb{R}\}$ are analytic solutions of an eigenvalue equation for $\hat{H}_r$; another aspect of $D_\lambda(z)$ is an analytic continuation of $\phi_{(n)}(x)$ with respect to $n$. The eigenstates $\{\phi_{(n)}(x),\bar{\phi}_{(n)}(x)\}$ and $\{D_\lambda(z),D_\rho(iz)\}$ stand on the same footing as scattering states of $\hat{H}_r$ unless any boundary conditions are added.

\section{Topics related to the present RHO formalism}

The complete  bases $\{\phi_\sigma\}$ or $\{\phi_{(n)},\bar{\phi}_{(n)}\}$ based on ladder operator $(A,\bar{A})$ give us useful ways to handle the problems related to RHOs; in what follows, we exhibit two simple examples. \vspace{3mm}

\subsection{Schwinger effect}

We note that the RHO is effectively realized by a particle interacting with a specific gauge field. Let us consider the scalar field $\Phi$ in four-dimensional spacetime for a mass $m$ particles under gauge fields $A^\mu$ satisfying
\footnote{
$\mbox{diag}(\eta_{\mu\nu})=(-+++)$.}
\begin{align}
 \left[ \hat{\Pi}_\mu(A)\hat{\Pi}^\mu(A)+(mc)^2\right]\Phi(x)=0, \label{KG-A}
\end{align}
where $\hat{\Pi}^\mu(A)=\hat{p}^\mu-\frac{g}{c}A^\mu$ and $g=\pm|e|$.  We here setup the gauge potentials in such a way that $(A_c^0(x),\bm{A}_c(x))=(-Ex^1, \bm{0})\,(E=\mbox{const}>0)$,  which produces the uniform electric field $E$ along the $x^1$ direction.  Then, 
\begin{align}
 \hat{\Pi}(A_c)^2=2m\hat{H}_{01}+\hat{p}_\perp^2, \label{PiPi}
\end{align}
where $\hat{p}_\perp=(\hat{p}_2,\hat{p}_3)$ and 
\begin{align}
 \hat{H}_{01} &=\frac{1}{2m}\hat{p}_1^2-\frac{1}{2m}\left(\frac{|e|E}{c}\right)^2\left(x^1+\frac{c}{gE}\hat{p}^0\right)^2. \label{H_01}
\end{align}
Further, in terms of the canonical variables defined by the unitary transformation $U_E=e^{\frac{i}{\hbar}\left(\frac{c}{gE}\right)\hat{p}^0\hat{p}^1}$ so that
\begin{align}
 (X^\mu) &=(U_Ex^\mu U_E^{-1})=\left(x^0-\frac{c}{gE}\hat{p}^1,x^1+\frac{c}{gE}\hat{p}^0,x^2,x^3 \right), \label{canonical-1}\\
 (\hat{P}^\mu) &=(U_E\hat{p}^\mu U_E^{-1})=(\hat{p}^\mu), \label{canonical-2}
 \end{align}
the Hamiltonian operator (\ref{H_01}) can be written as
\begin{align}
 \hat{H}_{01}=\frac{1}{2m}\hat{P}_1^2-\frac{m\omega^2}{2}X_1^2, \label{Hamiltonian_01}
\end{align}
where the angular frequency is defined by $\omega=\frac{|e|E}{mc}$. This means that the $H_{01}$ is just the Hamiltonian of the RHO defined in the phase space $(X^\mu,P^\mu)$.

Now, the classical action of the gauge field under consideration is $S_G[A_c]=\frac{1}{2}\int d^4x E^2$ and the one loop correction due to the scalar field $\Phi$ adds the quantum effect $S_{Q}[A_c]=-i\hbar\log\{\det(\Hat{\Pi}(A_c)^2+(mc)^2)\}^{-1}$ to $S_G[A_c]$
\footnote{In the expression of $S_Q[A_c]$, use has been made of  the well-known formulas 
$\{\det(M)\}^{-1}=e^{-\mbox{tr}\log M}$ and $\mbox{tr}\log M=-{\rm tr}\int_0^\infty\frac{d\tau}{\tau}e^{-i\tau(M-i\epsilon)}\,(+\mbox{const})$. }. The resultant effective action of gauge fields $S_{\rm eff}[A_c]=S_G[A_c]+S_Q[A_c]$ becomes
\begin{align}
 S_{\rm eff}[A_c] &=S_G[A_c] \nonumber \\
  &-i\hbar\int_0^\infty\frac{d\tau}{\tau}e^{-i\tau((mc)^2-i\epsilon)}\mbox{tr}\left(e^{-i\tau\hat{\Pi}[A_c]^2)}\right) \label{effective action}  
\end{align}
disregarding unimportant additional constant. Namely, under the classical background gauge field $A_c^\mu$, the scalar QED gives rise to the transition amplitude $\langle 0_{\rm in}|0_{\rm out}\rangle\sim Ne^{\frac{i}{\hbar}S_{\rm eff}[A_c]}\,(|N|^2=1)$, which defines an unitary $S$-matrix element for a real $S_{\rm eff}[A_c]$.  If the $S$ matrix contains pair productions, under which the state of the electric field is constant in time, then $\mbox{Im}S_{\rm eff}[A_c]\neq 0$ and we have 
 $|\langle 0_{\rm in}|0_{\rm out}\rangle|^2 \sim e^{-\frac{2}{\hbar}\mbox{Im}S_{\rm eff}[A_c]}\neq 1$. This ratio, the Schwinger effect, can be evaluated by calculating the \lq\lq trace\rq\rq \,in Eq.(\ref{effective action}).

 For this purpose, it is convenient to use $\{ |X^0\rangle\otimes |\phi_{(n)}(X^1)\rangle\otimes|X_\perp\rangle\}$ as the base states in the trace calculation. Then, by taking $1=\sum_{n=0}^\infty \frac{1}{N_n}|\phi_{(n)}\rangle\langle\bar{\phi}_{(n)}|$ and $\int dX^1 \bar{\phi}_{(n)}(X^1)^*\phi_{(n)}(X^1)=N_n$ into account, we obtain
\begin{align}
 \frac{1}{\hbar} &{\rm Im}S_{\rm eff}[A_c]=-{\rm Re} \int_0^\infty \frac{
d\tau}{\tau} e^{-i\tau((mc)^2-i\epsilon)} \nonumber \\
 &\times \mbox{tr}\left(e^{-i\tau\left(2m \hat{H}_{01}+\hat{P}_\perp^2\right) } \right)  \nonumber \\
 &=-{\rm Re}\int_0^\infty \frac{d\tau}{\tau} e^{-i\tau((mc)^2-i\epsilon)}\sum_{n=0}^\infty e^{-\tau2m\hbar\omega(n+\frac{1}{2})} \nonumber\\
 &\times \int dX^0\delta(0)\int d^2X_\perp \frac{\pi}{(2\pi\hbar)^2i\tau}~~\left(\,\delta(0)\sim \frac{\sqrt{m\hbar\omega}}{(2\pi\hbar)} \,\right), \label{ImS}
\end{align}
where the integral in $\delta(0)=(2\pi\hbar)^{-1}\int dP^0$ has been cut by the typical momentum scale in the $(X^0,X^1)$  so that $\int dP^0\sim\sqrt{m\hbar\omega}$ (Appendix C). Putting here $V_0\sim\int dX^0$, $V_\perp\sim\int d^2X_\perp$ as cutoff volumes in $X^0,X_\perp$ spaces respectively, the right-hand side of this equation becomes, with  $\epsilon=+0,$ 
\begin{align}
\mbox{rhs} 
 &=\delta(0)V_0V_\perp \mbox{Re}\left(\frac{i\pi}{(2\pi\hbar)^2}\int_{\epsilon}^\infty\frac{d\tau}{\tau^2}\frac{e^{-i\tau(mc)^2}}{2\sinh(\tau m\hbar\omega)} \right) \nonumber \\
 &=\delta(0) V_0V_\perp \frac{i\pi(mc)^2}{4(2\pi\hbar)^2} \mbox{P} \int_{-\infty}^\infty\frac{dz}{z^2}\frac{e^{-iz}}{\sinh \left(z\frac{\hbar\omega}{mc^2}\right)} \nonumber\\
 &=\delta(0) V_0V_\perp \frac{(m\hbar\omega)}{2(2\pi\hbar)^2} \sum_{n=1}^\infty \frac{(-1)^{n+1}}{n^2}e^{-\frac{n\pi c^2}{\hbar}\left(\frac{m^2c}{|e|E}\right)},  \label{ImS rhs}
\end{align}
where the $\mbox{P}$ denotes the principal value in the $z\,(=\tau(mc)^2)$ integral. In terms of  the momentum scale $\sqrt{m\hbar\omega}$, the volume $\int dX^1$ should also be given by\\
\begin{align}
 V_1\frac{\sqrt{m\hbar\omega}}{k\hbar}\sim 1~\left(V_1\sim\int dX^1\right) 
\end{align}
with a dimensionless constant $k$. Thus, with $V_{(4)}=V_0V_1V_\perp$ and $\omega=m\left(\frac{|e|E}{m^2c}\right)$, we arrive at the expression
\begin{align}
 \frac{1}{\hbar}\mbox{Im}S_{\rm eff}[A_c]\!\sim\! \frac{1}{k}\frac{V_{(4)}m^4}{16\pi^3\hbar^2}\!\left(\frac{|e|E}{m^2 c}\right)^2\!\sum_{n=1}^\infty \frac{(-1)^{n+1}}{n^2}e^{-\frac{n\pi c^2}{\hbar}\left(\frac{m^2c}{|e|E}\right)}. \label{S-effect}
\end{align}
For $k=1$, the result coincides with the formula of the pair creation given by Schwinger for scalar QED\cite{Schwinger,Dunne}. It should also be noticed that one can replace the constraints on the cutoff parameters by a weaker condition $\frac{(2\pi\hbar)\delta(0)}{m\omega V_1}\sim\frac{1}{k}$, which allows another possible choice such as $(2\pi\hbar)\delta(0)\sim\frac{\hbar\omega}{c}$ and $V_1\sim\frac{k\hbar}{mc}$.

\subsection{Extension to SUSY quantum mechanics}

The present ladder operator formalism of RHOs is easily extended to one of SUSY quantum mechanics\cite{SUSY-QM,Nicolai,Witten-1,Witten-2} . To this end, let us introduce the Fermi oscillators characterized by $\{b,b^\dag\}=1,\,b^2=b^{\dag 2}=0$, which can be represented in 2-dimensional vector space so that
\begin{align}
 b=\begin{pmatrix} 0 & 0 \\ 1 & 0 \end{pmatrix}\,, ~~~b^\dag=\begin{pmatrix} 0 & 1 \\ 0 & 0 \end{pmatrix}\,.
\end{align}
In terms of $(b,b^\dag)$,  the supersymmetric extension of $\hat{H}_r$ should be
\begin{align}
 \hat{\mathcal{H}}_r=-i\hbar\omega\left(-i\bar{A}A+b^{\dag} b \right)=-i\hbar\omega\begin{pmatrix} \bar{\Lambda} & 0 \\ 0 & \Lambda \end{pmatrix} .
\end{align}
Then the generators of SUSY transformation defined by
\begin{align}
\begin{split}
 Q &=-i\sqrt{\hbar\omega}Ab^\dag=\sqrt{\hbar\omega}\begin{pmatrix} 0 & -iA \\ 0 & 0 \end{pmatrix}, \\
 \bar{Q} &=-i\sqrt{\hbar\omega}\bar{A}b=\sqrt{\hbar\omega}\begin{pmatrix} 0 & 0 \\ -i\bar{A} & 0 \end{pmatrix}\,,
\end{split}
\end{align}
are characterized by the algebras
\begin{align}
\begin{split}
 [Q,A] &=\{Q,b^\dag\}=0, \\
 [Q,\bar{A}] &=\sqrt{\hbar\omega}b^\dag,~\{Q,b\}=-i\sqrt{\hbar\omega}A, \\
\end{split}
\end{align}
\begin{align}
\begin{split}
 [\bar{Q},\bar{A}] &=\{\bar{Q},b\}=0, \\
 [\bar{Q},A] &=-\sqrt{\hbar\omega}b,~\{\bar{Q},b^\dag\}=-i\sqrt{\hbar\omega}\bar{A},
\end{split}
\end{align}
and
\begin{align}
 [Q,\hat{\mathcal{H}}_r]=[\bar{Q},\hat{\mathcal{H}}_r]=0\,,~~\{Q,\bar{Q}\}=\hat{\mathcal{H}}_r \,.
\end{align}
If we introduce $Q_1=\frac{1}{\sqrt{2}}(\bar{Q}+Q)$ and $Q_2=\frac{i}{\sqrt{2}}(\bar{Q}-Q)$, the last equations can also be written as
\begin{align}
 \{Q_i,Q_j\}=\delta_{ij}\hat{\mathcal{H}}_r,~(i,j=1,2).
\end{align}
Those algebras should be compared with that of $N=2$ SUSY quantum mechanics, though $Q_i\,(i=1,2)$ are not Hermitian operators. The zero-point oscillation of $\hat{\mathcal{H}}_r$ is removed by this supersymmetry.

In spite of the formal resemblance of the present dynamical system to SUSY quantum mechanics of HOs, the true nature of both dynamical systems are fairly different as can be seen from $\bar{Q}\neq Q^\dag$, the nonpositive structure of $\hat{\mathcal{H}}_r$, and so on. On the discrete complete basis $\{\phi_{(n)},\bar{\phi}_{(n)}\}$, the eigenvalue equation   $\hat{\mathcal{H}}_r|\phi_E\rangle=E|\phi_E\rangle$ can be solved easily: for $n=1,2,\cdots$,
\begin{align}
 E^{+}_n &=-i\hbar\omega n \mbox{~~doublet} \\
 |\phi^{-}_n\rangle&=\begin{pmatrix} 0 \\ \phi_{(n)}\end{pmatrix} =|-\rangle\otimes|\phi_{(n)}\rangle \nonumber \\
 |\phi^{+}_n\rangle &=\frac{Q}{\sqrt{\hbar\omega n}}|\phi^{-}_n \rangle=\begin{pmatrix} -\frac{i}{\sqrt{n}}A\phi_{(n)} \\ 0 \end{pmatrix} \nonumber 
\end{align}
\begin{align}
  E^{-}_n &=i\hbar\omega n \mbox{~~doublet}  \\
|\bar{\phi}^{+}_n\rangle &=\begin{pmatrix} \bar{\phi}_{(n)} \\ 0\end{pmatrix} =|+\rangle\otimes|\bar{\phi}_{(n)}\rangle \nonumber \\
 |\bar{\phi}^{-}_n\rangle &=\frac{\bar{Q}}{\sqrt{\hbar\omega n}}|\bar{\phi}^{+}_{n}\rangle=\begin{pmatrix} 0 \\-\frac{i}{\sqrt{n}}\bar{A}\bar{\phi}_{(n)}\end{pmatrix} , \nonumber
\end{align}
where $|-\rangle=\begin{pmatrix}0\\1\end{pmatrix}$ and $|+\rangle=\begin{pmatrix}1\\0\end{pmatrix}$. Here, the mapping $|\phi^{+}_n\rangle=\frac{Q}{\sqrt{\hbar\omega n}}|\phi^{-}_n\rangle,(n\geq 1)$ can be inverted by $|\phi^{-}_n\rangle=i\frac{\bar{Q}}{\sqrt{\hbar\omega n}}|\phi^{+}_n\rangle$, and so the states $|\phi^{\pm}_n\rangle,\, (n=1,2,\cdots)$ form a tower of super pairs. In the same sense,  the states $|\bar{\phi}^{\pm}_n\rangle,\,(n=1,2,\cdots)$ form another tower of super pairs . 

\begin{figure}
\hspace{7mm}
\center
 \includegraphics[width=5.5cm]{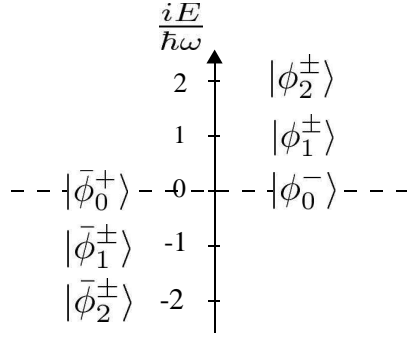}
\caption{The eigenvalues of $\hat{\mathcal{H}}_r$ in an extended SUSY quantum mechanics are illustrated. The states with the superscript $\pm$ are pair states mapped by $Q$ or $\bar{Q}$. The states $|\phi^{-}_0\rangle$ and $|\bar{\phi}^{+}_0\rangle$ are fixed states under those mapping.}
\label{SUSY eigenstates}
\end{figure}

In contrast, the states $|\phi^{-}_0\rangle$ and $|\bar{\phi}^{+}_0\rangle$ belonging to the same eigenvalue $E^{\pm}_0=0$ are two super singlets, which  satisfy $Q_i|\phi^{-}_0\rangle=Q_i|\bar{\phi}^{+}_0\rangle=0,(i=1,2)$ . Therefore in the space of states $\{|\phi^{-}_0\rangle,\{|\phi^{\pm}_n\rangle\}\}$, the supersymmetry is realized as a good symmetry; that is, SUSY is not broken. The same is true for the space of states $\{|\bar{\phi}^{+}_0\rangle,\{|\bar{\phi}^{\pm}_n\rangle\}\}$. In each space, the operators $Q_i$ work as the generators of supersymmetry; however, there arises no
mapping between those two spaces by $Q_i$ (Fig.\,\ref{SUSY eigenstates}).

In the context of this SUSY quantum mechanics, we emphasize the following: in the Schwinger effect for fermions, the SUSY quantum mechanics of the RHO plays an effective role in its background; that is, topics {III.A} and {III.B} are not independent in this effect. 

The Dirac field $\Psi$ interacting with an external gauge field $A^\mu$ obeys the $U(1)$ symmetry field equation
\footnote{The gamma matrices are normalized so that $\{\gamma^\mu,\gamma^\nu\}=-2\eta^{\mu\nu}$. }
\begin{align}
 \left(\gamma\cdot\hat{\Pi}(A)+mc\right)\Psi=0~~\,\left(\, \gamma\cdot\Pi(A)=\gamma_\mu\Pi(A)^\mu \,\right). \label{Dirac}
\end{align}
When we multiply this equation  by $- \left(\gamma\cdot\hat{\Pi}(A)-mc\right)$ from the left, the field equation becomes the second order form such that
\begin{align}
 \left[-\left(\gamma\cdot\hat{\Pi}(A)\right)^2+(mc)^2 \right]\Psi=0.
\end{align}
Here, if we use the configuration of gauge potentials $(A_c^0(x),\bm{A}_c(x))=(-Ex^1,\bm{0})$ as in Eq.(\ref{PiPi}),  then with $\sigma^{\mu\nu}=\frac{i}{2}[\gamma^\mu,\gamma^\nu]$ and $F_{\mu\nu}=\partial_{[\mu}(A_c)_{\nu]}$, we obtain
\begin{align}
 -\left(\gamma\cdot\hat{\Pi}(A_c)\right)^2 &=\hat{\Pi}(A_c)^2-\frac{\hbar g}{2c}\sigma_{\mu\nu}F^{\mu\nu} \nonumber \\
 &=\hat{\Pi}(A_c)^2-\frac{\hbar g}{c}\sigma^{01}E  \nonumber \\
 &=2m\hat{H}_{01}+\hat{p}_\perp^2-\frac{\hbar g}{c}(i\sigma_1\otimes\sigma_1)E. \label{gamma-Pi}
\end{align}
Carrying out the unitary transformation in 4-spinor space by $U=e^{\frac{i\pi}{4}\sigma_2}\otimes e^{\frac{i\pi}{4}\sigma_2}$, the Eq.(\ref{gamma-Pi}) becomes
\begin{align}
 -U &\left(\gamma\cdot\hat{\Pi}(A_c)\right)^2U^\dag \nonumber \\
 &=2m\left[-i\hbar\omega\left(-i\bar{A}A+\frac{1}{2}\right) \right]-m\hbar\omega i\sigma_3\otimes\sigma_3+\hat{p}_\perp^2 \nonumber \\
 &=2m\left[ -i\hbar\omega\begin{pmatrix} -i\bar{A}A+b^\dag b & 0 \\ 0 & -i\bar{A}A+bb^\dag \end{pmatrix} \right]+\hat{p}_\perp^2,
\end{align}
where we have used $\sigma_3\otimes\sigma_3=\sigma_3\otimes [b^\dag,b]$ and $\omega=\frac{|e|E}{mc}$ as before. The result implies that the spectra of upper components of $\tilde{\Psi}=U\Psi$ are those of the supersymmetric Hamiltonian $\hat{\mathcal{H}}_r$; on the other side, the spectra of lower components of $\tilde{\Psi}$ are governed by $\hat{\mathcal{H}}^\prime_r$, which is obtained from $\hat{\mathcal{H}}_r$ changing the role of $(b,b^\dag)$. Thus, one can evaluate the Schwinger effect for fermions again according to the procedure of Eqs. (\ref{ImS}) and Eq.(\ref{ImS rhs}) (Appendix D).

\section{Summary}

In this paper, we have discussed the eigenvalue problems of RHOs in terms of ladder operators $(A,\bar{A})$ introduced by an analogous way to the ladder operator $(\hat{a},\hat{a}^\dag)$ in HOs. The nonpositive property of the Hamiltonian operator $\hat{H}_r$ in RHOs is a result of the property of ladder operators such as $A^\dag=A$, $\bar{A}^\dag=\bar{A}$, and $[A,\bar{A}]=i$. Then, the eigenstates $A\phi_\sigma=\sigma\phi_\sigma$ and $\bar{A}\bar{\phi}_\sigma=\sigma\bar{\phi}_\sigma~(\sigma\in \mathbb{R})$ are able to normalize so that $\langle\phi_\sigma|\phi_{\sigma^\prime}\rangle=\langle\bar{\phi}_\sigma|\bar{\phi}_{\sigma^\prime}\rangle=\delta(\sigma-\sigma^\prime)$. Namely, the $\{\phi_\sigma,\bar{\phi}_\sigma\}$ are continual bases of the space of functionals $\bm{\Phi}^{\times}$ including the Hilbert space $\mathcal{H}$ of the RHO in the framework of rigged Hilbert space. Those continual bases are not eigenstates of $\hat{H}_r$, but rather the states related to $\{|x\rangle,\,|p\rangle \}$ by a unitary transformation.

On the other hand, the states $\phi_{(n)}=\bar{A}^n\phi_{(0)}$ and $\bar{\phi}_{(n)}=A^n\bar{\phi}_{(0)}$ $(n\in \mathbb{N})$ with $(\phi_{(0)},\bar{\phi}_{(0)})=(\phi_0,\bar{\phi}_0)$ are eigenstates of $\hat{H}_r$ belonging to the eigenvalues $\pm i\hbar\omega\left(n+\frac{1}{2}\right)$. Since those states satisfy the normalization of the form $\langle\bar{\phi}_{(m)}|\phi_{(n)}\rangle=\delta_{m,n}N_n$, it can be shown that the $\{\phi_{(n)},\bar{\phi}_{(n)}\}$ form a discrete complete basis of $\bm{\Phi}^{\times}$ in pairs. Contrary to this, the discrete eigenstates $\{\Phi_n\}$ of the Hamiltonian for a HO  are the basis of a Hilbert space. 

We can also show that Weber's $D$ functions, the special functions known as analytic solutions of the eigenvalue equation for $\hat{H}_r$ with continuous eigenvalues, are obtained by means of the analytic continuation of $\{\phi_{(n)},\bar{\phi}_{(n)}\}$ with respect to $n$. The $D$ functions and $\{\phi_{(n)},\bar{\phi}_{(n)}\}$ stand on the same footing as the scattering states of $\hat{H}_r$ unless any boundary conditions are added. 

As good applications of this ladder operator formalism, we have shown two topics: the Schwinger effect in scalar QED and an extension of RHO to SUSY quantum mechanics. In the first, the Hamiltonian of particles interacting with a constant electric field is  shown to be canonically equivalent to one of RHOs and so the knowledge of RHOs is useful to handle the problem of pair production by the electric field.  Indeed, it has been shown that the discrete complete bases $\{\phi_{(n)}, \bar{\phi}_{(n)}\}$ characterized by Eq.(\ref{unit operator}) give a simple way to evaluate such a production rate within the framework of quantum mechanics.

	Second, we have tried to extend the present RHO system to a supersymmetric dynamical system; the extended Hamiltonian $\hat{\mathcal{H}}_r$ is again a nonpositive Hermitian operator constructed out of fermionic  oscillators $(b,b^\dag)$ and ladder operators $(A,\bar{A})$. The ladder operator formalism gives rise to two towers of super-pair states $|\phi^{\pm}_n\rangle$ and $|\bar{\phi}^{\pm}_n\rangle$ $(n=1,2,\cdots)$, which belong to the eigenvalues $E^{+}_n=-i\hbar\omega n$ and $E^{-}_n=i\hbar\omega n$ respectively. In addition to this, the $n=0$ states $|\phi^{-}_0\rangle$ and $|\bar{\phi}^{+}_0\rangle$ exist as two singlet states, which satisfy  $Q_i|\phi^{-}_0\rangle=Q_i|\bar{\phi}^{+}_0\rangle=0,(i=1,2)$. Namely, in each space of super-pair tower states,  SUSY is realized as a good symmetry, though the SUSY in this model is an extended concept from the standard one as can be seen from $Q_i^\dag\neq Q_i$.

 Furthermore, we have brought up the following: if we consider the Dirac fields interacting with an external electric field, then the supersymmetric structure of RHOs will be implicitly included in a loop effect of those Dirac fields.  According to this line of approach, we have shown the way to evaluate the Schwinger effect for fermions in Appendix D.

The knowledge on the complete bases in RHOs under the ladder operator formalism is expected to give useful tools in various problems other than the topics discussed in this paper. For example, the Hamiltonian $\hat{\mathcal{H}}_r$ is able to take continuous eigenvalues on the states $(\phi_\sigma,\bar{\phi}_\sigma)$; in the space of those eigenstates, the SUSY may show a different feature from the standard analysis. Those are interesting future problems.

\section*{Acknowledgments}

The authors wish to thank the members of the theoretical group at Nihon University for their hospitality. The authors also appreciate one of the referees concerning improvements to the descriptions of Sec. II.

\appendix

\section{WEBER'S FUNCTIONS AS THE ENERGY EIGENVALUE FUNCTIONS FOR THE RHO}

We here summarize the standard way to make the eigenvalue functions of the RHO reduce to Weber's functions. 

In the $x$ representation with $\hat{p}=-i\hbar\frac{\partial}{\partial x}$, the eigenvalue equation of $\hat{H}_r$ can be written as
\begin{align}
 \left(-\frac{\hbar^2}{2m}\frac{d^2}{dx^2}-\frac{m\omega^2}{2}x^2-E\right)\psi_E(x)=0 .\label{eigen-1}
\end{align}
Introducing here the variable $z$ defined by
\begin{align}
 x=e^{\frac{i\pi}{4}}\sqrt{\frac{\hbar}{2m\omega}}z\,,
~~~\left( \, \frac{d^2}{dx^2}=\frac{2m\omega}{i\hbar}\frac{d^2}{dz^2} \, \right),
\end{align}
Eq.(\ref{eigen-1}) with $\psi_E(x(z))=w_E(z)$ gives rise to
\begin{align}
 -\frac{i}{\hbar\omega}\times\mbox{Eq.}(\ref{eigen-1}) = \left(\frac{d^2}{dz^2}+\frac{iE}{\hbar\omega}-\frac{1}{4}z^2\right)w_E(z)=0 . \label{eigen-2}
\end{align}
Writing $\frac{iE}{\hbar\omega}=\lambda+\frac{1}{2}$ and $w_E(z)=w_{\lambda}(z)$,  Eq.(\ref{eigen-2}) becomes the standard form of Weber's equation
\begin{align}
\frac{d^2w_{\lambda}(z)}{dz^2}+\left(\lambda+\frac{1}{2}-\frac{z^2}{4}\right)w_{\lambda}(z)=0 . \label{Weber-eq}
\end{align}
For $\tilde{w}_\lambda(z)=e^{\frac{1}{4}z^2}w_\lambda(z)$, Eq.(\ref{Weber-eq}) can also be written as  
\begin{align}
 \left(\frac{d^2}{dz^2}-z\frac{d}{dz}+\lambda\right)\tilde{w}_\lambda(z)=0. \label{Weber-eq-2}
\end{align}

\begin{figure}\centering
 \includegraphics[width=6cm]{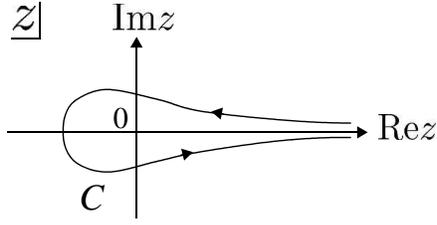}
\caption{Contour $C$ in a complex plane}
\label{contour}
\end{figure}

To solve Eq.(\ref{Weber-eq-2}), let us use  the Fourier-Laplace representation
\begin{align}
 \tilde{w}_\lambda(z)=\int_\Gamma dt e^{-zt}f_\lambda(t),
\end{align}where $\Gamma$ is a path from $a$ to $b$ in the complex $t$ plane. Then under the integration by parts with respect to $t$, Eq.(\ref{Weber-eq-2}) with Eq.(\ref{Weber-eq}) gives
\begin{align}
 \frac{d}{dt}\big\{tf_\lambda(t)\big\}+\left(t+\frac{\lambda}{t}\right)\big\{tf_\lambda(t)\big\}=0 \label{Weber-eq-3}
\end{align}
on the condition that $\left[e^{-zt}\left\{tf_\lambda(t)\right\}\right]^b_a=0$. Equation (\ref{Weber-eq-3}) can be solved easily so that $f_\lambda(t)=\mbox{const}\,e^{-\frac{1}{2}t^2}t^{-(\lambda+1)}$; since the boundary conditions are satisfied by $(a,b)=(0,\infty)$ for $\mbox{\rm Re}\lambda<0$ on the real $t$  axis,  and we finally obtain the integral representation for $w_\lambda(z)=e^{-\frac{1}{4}z^2}\tilde{w}_\lambda(z)\,(=D_\lambda(z))$ in such a form as
\cite{Bateman, Handbook}
\begin{align} 
 D_\lambda(z) &=\frac{e^{-\frac{1}{4}z^2}}{\Gamma(-\lambda)}\int_0^\infty dt
e^{-zt-\frac{1}{2}t^2}t^{-(\lambda+1)}\hspace{7mm}(\,\mbox{Re}\lambda < 0\,) \label{A-D-function}\\
 &=-\frac{\Gamma(\lambda+1)}{2\pi i}e^{-\frac{1}{4}z^2}\int_{C}dte^{-zt-\frac{1}{2}t^2}(-t)^{-(\lambda+1)}, \label{A-D-function-2}
\end{align}
where $C$ is the contour given in Fig.\,\ref{contour}. It is not difficult to rewrite the contour integral in Eq.(\ref{A-D-function-2}) to the path integral in Eq.(\ref{A-D-function}) by taking into account $\Gamma(\lambda+1)\sin(-\pi\lambda)=\frac{\pi}{\Gamma(-\lambda)}$.

The function $D_\lambda(z)$ is Weber's $D$-function
\footnote{The $D$ function is normalized so that $D_n(z),\,(n=0,1,\cdots)$ reduces to $e^{-\frac{1}{4}z^2}H_{e_n}(z)$, where $\{H_{e_n}(z)\}$ are the Chebyshev-Hermite polynomials.  }
 (Parabolic cylinder function)\cite{Handbook}, by which the independent solutions of Eq.(\ref{eigen-1}) for $\frac{iE}{\hbar\omega}=\lambda+\frac{1}{2}$ are given as $D_\lambda(z)$ and $D_{-\lambda-1}(iz)$.

\section{ANOTHER PROOF OF $\hat{I}_{r}=1$.}

By the definitions of $\phi_{(n)}$ and $\bar{\phi}_{(n)}$, we obtain the expression
\begin{align} 
\langle x|\hat{I}_{r}|x^\prime\rangle &=\sum_{n=0}^\infty\frac{1}{N_n}\phi_{(n)}(x)\bar{\phi}_{(n)}(x^\prime)^* \nonumber \\
 =\sqrt{\frac{2\pi}{i}} &\sqrt{\frac{m\omega}{2\hbar\pi^2}}\sum_{n=0}^\infty \frac{1}{n!}\left(e^{-\frac{i\pi}{2}}\bar{A}A^{\prime *}\right)^n
e^{i\frac{m\omega}{2\hbar}x^2}e^{i\frac{m\omega}{2\hbar}x^{\prime 2}} \nonumber \\
 =\sqrt{\frac{2\pi}{i}} &\sqrt{\frac{m\omega}{2\hbar\pi^2}}e^{e^{-\frac{i\pi}{2}}\bar{A}A^{\prime*}}e^{i\frac{m\omega}{2\hbar}x^2}e^{i\frac{m\omega}{2\hbar}x^{\prime 2}} \nonumber \\
 =\sqrt{\frac{2\pi}{i}} &\sqrt{\frac{m\omega}{2\hbar\pi^2}}e^{i\frac{m\omega}{2\hbar}x^2}e^{-2iA^{\prime*}\sqrt{\frac{m\omega}{2\hbar}}x}e^{\frac{i}{2}\left(A^{\prime*}\right)^2}e^{i\frac{m\omega}{2\hbar}x^{\prime 2}}.
\end{align} 
Here, we have used the formula $e^{\hat{a}+\hat{b}}=e^{\hat{a}}e^{\hat{b}}e^{-\frac{1}{2}[\hat{a},\hat{b}]}$ for $[[\hat{a},\hat{b}],\hat{a}]=[[\hat{a},\hat{b}],\hat{b}]=0$, with $\hat{a}=e^{-\frac{i\pi}{2}}A^{\prime*}\sqrt{\frac{m\omega}{2\hbar}}x$ and $\hat{b}=e^{-\frac{i\pi}{2}}A^{\prime*}\frac{-1}{\sqrt{2m\hbar\omega}}\hat{p}$. Remembering, further, 
\begin{align}
e^{\frac{i}{2}(A^{\prime*})^2}=\sqrt{\frac{i}{2\pi}}\int_{-\infty}^\infty dk
e^{-\frac{i}{2}k^2+ikA^{\prime*}}
\end{align}
and using again $e^{\hat{a}+\hat{b}}=e^{\hat{a}}e^{\hat{b}}e^{-\frac{1}{2}[\hat{a},\hat{b}]}$, we arrive at 
\begin{align}
\langle x|\hat{I}_{r}|x^\prime\rangle &=\sqrt{\frac{m\omega}{2\hbar\pi^2}}e^{i\frac{m\omega}{2\hbar}x^2}\int_{-\infty}^\infty dk  \nonumber \\
 &\times e^{-\frac{i}{2}k^2}e^{-i\left(\sqrt{\frac{2m\omega}{\hbar}}x-k\right)A^{\prime*}}e^{i\frac{m\omega}{2\hbar}x^{\prime 2}} \nonumber \\
 &=\sqrt{\frac{m\omega}{2\hbar\pi^2}}e^{i\frac{m\omega}{2\hbar}x^2}\int_{-\infty}^\infty dk
 \nonumber \\
 &\times e^{-\frac{i}{2}k^2}e^{-i\left(\sqrt{\frac{2m\omega}{\hbar}}x-k\right)\sqrt{\frac{m\omega}{2\hbar}}x^\prime}e^{\frac{i}{4}\left(\sqrt{\frac{2m\omega}{\hbar}}x-k\right)^2} \nonumber \\
 &\times e^{i\frac{m\omega}{2\hbar}\left\{x^\prime-\left(\sqrt{\frac{2m\omega}{\hbar}}x-k\right)\sqrt{\frac{\hbar}{2m\omega}}\right\}^2} \nonumber \\
 &=\sqrt{\frac{m\omega}{2\hbar\pi^2}}e^{i\frac{m\omega}{2\hbar}x^2}\times e^{-i\frac{2m\omega}{\hbar}xx^\prime+i\frac{m\omega}{\hbar}x^2+i\frac{m\omega}{2\hbar}x^{\prime 2}} \nonumber \\
 &\times  \int_{-\infty}^\infty dk e^{ik\sqrt{\frac{2m\omega}{\hbar}}(x-x^\prime)}=\delta(x-x^\prime). 
\end{align} 
Therefore, $\hat{I}_{r}$ is nothing but the unit operator for the present RHO system. 

\section{TYPICAL MOMENTUM SCALE IN $(X^0,X^1)$ SPACE} 

The canonical pairs $(\hat{X}^\mu,\hat{P}^\mu),\,(\mu=0,1)$ in Eqs. (\ref{canonical-1})-(\ref{canonical-2}) can be equivalently represented by the pairs $(A,\bar{A}^\dag)$ and $(B,B^\dag)$ defined by
\begin{align}
 (\hat{X}^1,\hat{P}^1) &=\left(\sqrt{\frac{\hbar}{2m\omega}}(\bar{A}-A),\sqrt{\frac{m\hbar\omega}{2}}(\bar{A}-A)\right) \\
(\hat{X}^0,\hat{P}^0) &=\left(\sqrt{\frac{\hbar}{2m\omega}}(B^\dag+B),i\sqrt{\frac{m\hbar\omega}{2}}(B^\dag-B)\right).
\end{align}
Under the canonical commutation relations $[\hat{X}^\mu,\hat{P}^\nu]=i\hbar\eta^{\mu\nu}$, the $(A,\bar{A}^\dag)$ become the ladder operators for a RHO, and  the $(B,B^\dag)$ are oscillator variables for a HO characterized by $[B,B^\dag]=1$. 

In classical, the momentum square in the phase space $(X^1,P^1)$ can be identified with the radius square $R_1^2=2mH_{01}=-m\hbar\omega\left(\bar{A}A+A\bar{A}\right)$, where ${H}_{01}$ is the classical counterpart of the Hamiltonian (\ref{Hamiltonian_01}). Similarly, the momentum square in the phase space $(X^0,P^0)$ should be given by $R_0^2=m\hbar\omega\left(B^\dag B+BB^\dag\right)$. In q-number theory, those radii should be evaluated by means of the expectation values through the partition functions
\begin{align}
 Z_{\mu}[\epsilon]={\rm Tr}e^{-i^{\delta_{1,\mu}}\epsilon\hat{R}_\mu^2}~~(\mu=0,1),
 \end{align}
where the $\epsilon$ is a regularization parameter for those partition functions. It should be noticed that the $\hat{R}_0$ and the $i\hat{R}_1$ have real positive eigenvalues on their complete basis $\left\{|n\rangle_B=(n!)^{-\frac{1}{2}}B^{\dag n}|0\rangle\right\}$ and $\left\{|\phi_n\rangle=\bar{A}^n|\phi_{(0)}\rangle\right\}$ with $B|0\rangle=0\,(\langle 0|0\rangle=1)$ and $A|\phi_{(0)}\rangle=\bar{A}|\bar{\phi}_{(0)}\rangle=0\,\left(\langle\bar{\phi}_{(0)}|\phi_{(0)}\rangle=\sqrt{\frac{i}{2\pi}}\right)$. Then the expectation values $\langle R^2_\mu\rangle$ can be evaluated so that
\begin{align}
 \langle R^2_\mu\rangle=-\frac{\partial}{\partial\epsilon}\log Z_\mu[\epsilon]=\hbar m\omega+\frac{2m\hbar\omega}{e^{\epsilon2m\hbar\omega}-1}. \label{radius square}
 \end{align}
 The first term $\hbar m\omega$ in the rightest side of Eq.(\ref{radius square}) represents a typical momentum square in $(X^0,X^1)$ space, which is independent of the regularization.

\section{THE SCHWINGER EFFECT FOR FERMIONS}

The action of the Dirac field $\Psi$ obeying Eq.(\ref{Dirac}) with the gauge fields $A_c^\mu$ is $S_D[\Psi,A_c]=\int d^4 x\bar{\Psi}\left(\gamma\cdot\hat{\Pi}(A_c)+mc\right)\Psi$,\, $\left(\bar{\Psi}=\Psi^\dag\gamma^0\right)$. Then the path integral result $S_Q[A_c]=-i\hbar\log\int\mathcal{D}\Psi\mathcal{D}\bar{\Psi}e^{\frac{i}{\hbar}S_D}=-i\hbar\mbox{Tr}\log(\gamma\cdot\hat{\Pi}+mc)+\mbox{const}$ is the quantum correction to $S_G[A_c]=\frac{1}{2}\int d^4x E^2$ so that $S_{\rm eff}[A_c]=S_G[A_c]+S_Q[A_c]$ becomes the effective action of $A_c$. Here, the \lq\lq$\mbox{Tr}$\rq\rq ~ involves the trace over four-component spinor space. To evaluate the $\mbox{Tr}$  in $S_Q[A_c]$, we notice that
\begin{align}
 \frac{d}{da}\mbox{Tr} &\log\Big[\gamma\cdot\hat{\Pi}+(mc)+a\Big] \!=\!\mbox{Tr}\frac{-(\gamma\cdot\hat{\Pi})+\{(mc)+a\}}{-(\gamma\cdot\hat{\Pi})^2+\{(mc)+a\}^2} \nonumber \\
 &=\mbox{Tr}\frac{\{(mc)+a\}}{-(\gamma\cdot\hat{\Pi})^2+\{(mc)+a\}^2} \nonumber \\
 &=\frac{d}{da}\left(-\frac{1}{2}\mbox{Tr}\int_0^\infty\frac{d\tau}{\tau}e^{-i\tau\left[-(\gamma\cdot\hat{\Pi})^2+\{(mc)+a\}^2 -i\epsilon\right] }\right),
\end{align}
in consideration of which the trace of odd powers of $\gamma$ matrices vanishes. Integrating this equation with respect to $a$ from $a_1$ to $a_2$, we obtain
\begin{align}
 \mbox{Tr} &\log\left[\gamma\cdot\hat{\Pi}+(mc)+a_2\right] -\mbox{Tr}\log\left[\gamma\cdot\hat{\Pi}+(mc)+a_1\right]  \nonumber \\
 &=-\frac{1}{2}\mbox{Tr}\int_0^\infty\frac{d\tau}{\tau}e^{i\tau\{(\gamma\cdot\hat{\Pi})^2+i\epsilon\}} \nonumber \\
 &\times \left[e^{-i\tau\{(mc)+a_2\}^2}-e^{-i\tau\{(mc)+a_1\}^2} \right] .
\end{align}
Setting $a_2=0$ and $a_1=-(\gamma\cdot\hat{\Pi}+mc)+1$, we get the expression 
\begin{align}
 \mbox{Tr}\log(\gamma\cdot\hat{\Pi}+mc)=-\frac{1}{2}\mbox{Tr}\int_0^\infty\frac{d\tau}{\tau}e^{i\tau(\gamma\cdot\hat{\Pi})^2}e^{-i\tau\{(mc)^2-i\epsilon\}} \label{Tr}
\end{align}
disregarding an unimportant additional constant. Then remembering Eq.(\ref{gamma-Pi}) and using $e^{i\tau\frac{\hbar g}{c}E\sigma^{01}}=\cosh\left(\tau\frac{\hbar gE}{c}\right)+i\sigma^{01}\sinh\left(\tau\frac{\hbar gE}{c}\right)$, Eq.(\ref{Tr}) becomes 
\begin{align}
  \mbox{Tr}\log(\gamma\cdot\hat{\Pi} &+mc)=-\frac{1}{2}\times 4\mbox{tr}\int_0^\infty\frac{d\tau}{\tau} \nonumber \\
 &\times e^{-i\tau\left(\hat{\Pi}^2+(mc)^2-i\epsilon\right)}\cosh\left(\tau\frac{\hbar gE}{c}\right) 
\end{align}
by virtue of the trace of $\sigma^{01}$ vanishes. Here, the \lq\lq$\mbox{tr}$\rq\rq\, denotes the trace in the functional space, which yields $\mbox{tr}e^{-i\tau\hat{\Pi}^2}=\delta(0)V_0V_\perp\left(\frac{1}{4\pi i\hbar^2\tau}\right)\frac{1}{2\sinh(\tau m\hbar\omega)}$ with $\omega=\frac{|e|E}{mc}$ and $\delta(0)=\frac{\sqrt{m\hbar\omega}}{(2\pi\hbar)}$ as in the case of scalar QED. Therefore, we arrive at the expression with $\epsilon=+0$
\begin{align}
  \mbox{Tr}\log(\gamma\cdot\hat{\Pi}+mc) &=-\delta(0)V_0V_\perp\nonumber \\
 \times \frac{2(mc)^2}{(2\pi\hbar)^2} &\frac{\pi}{i} \int_{\epsilon}^\infty\frac{dz}{z^2}\frac{e^{-iz}}{\tanh\left(z\frac{\hbar\omega}{mc^2\pi}\right)} , \label{tanh}
\end{align}
where $z=\tau(mc)^2$. The integration with respect to $z$ in Eq.(\ref{tanh}) can be carried out in the same manner as Eq.(\ref{ImS rhs}) except replacing the residue $(-1)^n$ of $1/\sinh(z\hbar\omega/mc^2)$ by $1$ of $1/\tanh(z\hbar\omega/mc^2)$. Using further $V_1\frac{\sqrt{m\hbar\omega}}{k\hbar}\sim 1$ with $k=1$, the resultant formula corresponding to Eq.(\ref{S-effect}) in the case of Dirac fields becomes
\begin{align}
 \frac{1}{\hbar}\mbox{Im}S_{\rm eff}[A_c] &=-\mbox{Re}\mbox{Tr}\log\left[\gamma\cdot\hat{\Pi}(A_c)+mc\right] \nonumber \\
 &\sim V_{(4)}\frac{m^4}{8\pi^3\hbar^2}\left(\frac{|e|E}{m^2c}\right)^2\sum_{n=1}^\infty\frac{1}{n^2}e^{-\frac{n\pi c^2}{\hbar}\left(\frac{m^2c}{|e|E}\right)}.
\end{align}
This formula is nothing but the one given originally by Schwinger\cite{Schwinger}.

 \vspace{10mm}

\end{document}